\documentclass[prb,preprint]{revtex4}
\usepackage{color} 
\usepackage{amssymb}
\usepackage{bm}
\newcommand{\be}{\begin{equation}}
\newcommand{\ee}{\end{equation}}
\newcommand{\bea}{\begin{eqnarray}}
\newcommand{\eea}{\end{eqnarray}}
\newcommand{\nn}{\nonumber}
\newcommand{\gp}{\left( \gamma\cdot \Pi \right)}
\newcommand{\Ep}{\mathbb{E}_p}

\newcommand{\Pit}{\mathbf{\tilde{\Pi}}}
\newcommand{\gammaPi}{\boldsymbol{\gamma \cdot  \Pi}}

\begin{document}

\title{Free Form of the Foldy-Wouthuysen Transformation in External Electromagnetic Fields}

\author{Gabriela Murgu\'{\i}a}
\email{murguia@ciencias.unam.mx}
\affiliation{Departamento de F\'{\i}sica, Facultad de Ciencias, Universidad Nacional Aut\'onoma de M\'exico. Apartado Postal 21-092, M\'exico, D. F., 04021, M\'exico}

\author{Alfredo Raya}
\email{raya@ifm.umich.mx}
\affiliation{Instituto de F\'{\i}sica y Matem\'aticas, Universidad Michoacana de San Nicol\'as de Hidalgo. Edificio C-3, Ciudad Universitaria, 58040, Morelia, Michoac\'an, M\'exico.}

\begin{abstract}
We derive the exact Foldy-Wouthuysen transformation for Dirac fermions in a time independent external electromagnetic field in the basis of the Ritus eigenfunctions, namely the eigenfunctions of the operator $\gp^2$, with $\Pi^\mu=p^\mu-eA^\mu$. In this basis, the transformation acquires a free form involving the dynamical quantum numbers induced by the field. 
\end{abstract}



\maketitle


The study of semiclassical and nonrelativistic limits of the Dirac equation is a useful method to understand some effects on fermions coupled to external fields. In both, gravitational~\cite{Obukhov:2009,Goncalves:2007} and electromagnetic backgrounds~\cite{Silenko:2008, Barducci:2009}, the Foldy-Wouthuysen (FW)  transformation~\cite{FW} has proven to be a favorite way to obtain the nonrelativistic limit of the Dirac equation, because it provides a block diagonal form representation of quantum operators and hence of the Dirac Hamiltonian itself. 
Powerful applications of FW transformation in semiclassical calculations can be exploited in systems of other than (3+1) dimensions~\cite{Matias:1992} due to its relation with the supersymmetric character, in the quantum mechanical sense~\cite{Cooper:art, Cooper}, of some class of Hamiltonians, as well as in different stationary metrics~\cite{Heidenreich:2006, Buhl:2008}.


In the free Dirac equation, 
\be
i \frac{\partial \Psi}{\partial t} = (\mathbf{\boldsymbol{\alpha} \cdot p} + \beta m ) \Psi ,
\label{Dirac}
\ee 
expressed in natural units, $\hbar = c =1$, with the matrices $\alpha^i$ and $\beta$ given in terms of the Dirac $\gamma^\mu$-matrices as $\alpha^i = \gamma^0 \gamma^i$ and $\beta = \gamma^0$,
the large and small components of the Dirac spinor $\Psi$, labeled by the momentum $\mathbf {p}$,  get mixed by the odd operator $\mathbf{\boldsymbol{\alpha} \cdot p}$  involving off-diagonal elements. The FW is a canonical transformation which,  by removing such an operator from the Dirac Hamiltonian, 
\be
H_{\rm free}=\mathbf{\boldsymbol{\alpha} \cdot p} + \beta m, 
\ee
allows to decouple these large and small components of  $\Psi$.
The free FW transformation, 
\be
e^{i S_{\rm free}(\mathbf{p})} = \cos{|\mathbf{p}|\theta} + 
                \frac{ \mathbf{\boldsymbol{\gamma} \cdot p}}{|\mathbf{p}|} \sin{|\mathbf{p}|\theta},
\ee
with $\theta$ given through
\be
\tan(2 |\mathbf{p}| \theta) = \frac{|\mathbf{p}|}{m} ,
\ee 
is exact, and renders the free Hamiltonian in the form
\be 
H_{\rm free}^{FW} =\gamma^0 \sqrt{\mathbf{p}^2 + m^2} .
\label{H_FW-free}
\ee 

In the presence of an external electromagnetic field, which enters in the Dirac Hamiltonian through the replacement $p^\mu\to \Pi^\mu=p^\mu - eA^\mu$ with $A^\mu=(\phi,\mathbf{A})$,  the FW transformation 
can be ordinarily obtained through successive approximations as an expansion in powers of $1/m$~\cite{BD}.
At $O(1/m^3)$, 
\be 
S(\Pi) = -i\left(\frac{\gamma^0}{2m}\right)^2 \left([\mathcal{O}^\prime,\mathcal{E}^\prime]+
                                            i \dot{\mathcal{O}}^\prime\right) ,
\ee 
with
\bea
\mathcal{O}^\prime &= \frac{\gamma^0}{2m}[\mathcal{O},\mathcal{E}] -
                                  \frac{\mathcal{O}^3}{3m^2} +
                                  i\frac{\gamma^0 \dot{\mathcal{O}}}{2m} ,\\
\mathcal{E}^\prime &= \mathcal{E} + 
                                  \gamma^0 \left( \frac{\mathcal{O}}{2m} - \frac{\mathcal{O}^4}{8m^3}\right)
                                  - \frac{1}{8m^2}[\mathcal{O},[\mathcal{O},\mathcal{E}]]
                                  - i\frac{1}{8m^2}[\mathcal{O},\dot{\mathcal{O}}] .
\eea 
In the expressions above, the even (diagonal) and odd (off-diagonal) operators correspond to $\mathcal{E} =  e\phi$ and $\mathcal{O} = \gamma^0\gammaPi \equiv \Pit$, respectively, and the dot represents the time derivative.
To this order, the FW transformation renders the Dirac Hamiltonian to its leading non-relativistic form.
 For an external static inhomogeneous magnetic field the series can be written in closed form as
\be 
H_{\Pi}^{FW} =\gamma^0 \sqrt{(\Pit)^2 + m^2} ,
\label{H_FW-Serie}
\ee 
where the transformation matrix for this case is
\be
S(\Pi) = -i (\gammaPi) \theta ,
\ee
with $\theta$ given through
\be
\tan(2 |\Pit| \theta) = \frac{|\Pit|}{m} .
\ee 
Here, $|\Pit| = \sqrt{(\gamma^0 \gammaPi)^2}$
plays the role of the momentum $|\mathbf{p}|$ in the free case.

In this article, we exploit a formalism,  the so-called Ritus  method~\cite{Ritus:1972, Ritus:1974, Ritus:1978},  to  write the exact FW transformation for a static external 
field in a closed form as well, but obtained more directly.


Ritus method is an alternative approach to deal with quantum field theory in background electromagnetic fields. It was first developed to diagonalize, in momentum space, the spin-1/2 fermion propagator in the presence of a uniform magnetic field~\cite{Ritus:1972, Ritus:1974, Ritus:1978}, and was later extended for the propagator of charged gauge bosons~\cite{cubanos:2002,cubanos:2004}. The generalization of the method for Dirac particles in non-uniform magnetic fields has recently been developed~\cite{edward}.  For a pedagogical introduction to Ritus formalism, see Ref.~\cite{GM-AR-AS-ER}.
The Ritus eigenfunctions $\Ep$ 
are the matrices that diagonalize the  operator $\gp^2$, 
\be
\Ep^{-1}\gp^2 \Ep = p^2,
\label{EigenvaluesEp2}
\ee 
and correspond to the asymptotic states of charged particles in the presence of external fields.
For Dirac fermions, the matrix form of the $\Ep$ functions is inherited from  their spin realization.  
These, however, cannot be expressed in a closed form for a general configuration of external fields. Fortunately,  in the case of a static inhomogeneous field directed, for instance, along the $z-$axis, in the gauge $A^\mu=(0,0,W(x),0)$ such that $B(x)=\partial_x W(x) \equiv W'(x)$, $\Ep$ functions can be expanded over an orthogonal set of functions provided $W(x)$ belongs to the class of shape invariant potentials in the supersymmetric quantum mechanical framework~\cite{Cooper:art, Cooper}. Similar arguments apply if we consider a static inhomogeneous electric field of the form $E(x)=-\phi'(x)$ with $A^0=\phi(x)$ in the above $A^\mu$.
For these field configurations of our interest, the $\Ep$ form a complete set 
\bea
\nn
\int d^4z \ \overline {\mathbb {E}}_{p'}(z) \mathbb {E}_p(z) = \hat{\delta}^{(4)}(p-p')\Pi(n), \\
{\int\!\!\!\!\!\!\!\!\sum} \frac{d^4p}{(2\pi)^4} \ \mathbb {E}_{p}(z) \overline {\mathbb {E}}_p (z') = \delta^{(4)}(z-z'), \label{ortonormalidad}
\eea
where $z=(x_0, x_1, x_2, x_3) = (t,x,y,z)$, $ \overline {\mathbb{E}}_p = \gamma^0 \mathbb {E}_p^\dagger \gamma^0 $,
and $\Pi(n)$ is a spin projector which singles out only one spin orientation in the lowest Landau level. The symbol $\int \!\!\!\!\!\!\sum d^4p$ indicates that the integration might represent a sum, depending upon the continuous or discrete nature of the components of  the momentum $p$, and correspondingly, $\hat\delta^{(4)}$ represents a product of Kronecker and Dirac delta functions.
 Ritus functions play the same role plane-waves do in absence of external fields, namely, these serve to expand the propagator in momentum space as
\be
S_F(z,z')= {\int\!\!\!\!\!\!\!\!\sum} d^3p \ d^3p' \ {\mathbb E}_p(z) S_F(p,p') \overline{\mathbb E}_{p'}(z')\;. \label{integ}
\ee
Here, the propagator $S_F(p,p')$ takes the form ~\cite{Ritus:1972, Ritus:1974, Ritus:1978, GM-AR-AS-ER} 
\be
S_F(p,p')=\hat{\delta}^{(4)}(p-p')\Pi(n)\tilde{S}_F(\overline{p}),
\ee
where
\be
\tilde{S}_F(\overline{p})=\frac{1}{\gamma\cdot \overline{p}-m}, 
\label{propagador_libre}
\ee
and the momentum $\overline{p}$ involves the dynamical quantum numbers induced by the external field with the property $\overline{p}^2=p^2$. This vector is usually defined through the gauge invariant relation
\be 
\gp\Ep = \Ep (\gamma\cdot \overline{p}), \label{gpEp}
\ee
although its explicit form depends on the external field under consideration. In the $\Ep$ basis, the Dirac wave function is expressed as
\be 
\Psi = \Ep u_{\overline{p}},\label{psi}
\label{DiracSpinor}
\ee 
and from the property~(\ref{gpEp}), it is easy to see that $ u_{\overline{p}}$ is a free spinor labeled by the momentum $\overline{p}$. With this decomposition of $\Psi$, the information of the interactions with the external field is factorized
into the ${\mathbb E}_p$ functions and the $\overline{p}$ dependence of $u_{\overline{p}}$. This can be implemented to derive a closed form for the FW transformation.

With the decomposition of the Dirac wave function (\ref{psi}), the stationary Schr\"odinger form of the Dirac equation becomes
\be
E_D  \Ep u_{\overline {p}} = \gamma^0 (\gammaPi + m) 
                                                       \Ep u_{\overline {p}}
\label{DiracEp}
\ee
which with the aid of property~(\ref{gpEp}), simplifies to
\be
E_D \Ep u_{\overline {p}} = 
  \Ep \gamma^0 (\boldsymbol{\gamma \cdot \overline{p}} + m) u_{\overline {p}} .
\label{DiracEpFree}
\ee
In the above expressions $E_D$ represent the eigenenergies of the Dirac equation. Moreover, the Hamiltonian on the r.h.s. of Eq.~(\ref{DiracEpFree}) acquires a free form involving $\overline{p}$ alone. Thus, it is straightforward to convince oneself that the Ritus eigenfunctions map the FW transformation in external fields to a free transformation which depends on $\overline{p}$, namely
\be
 e^{iS(\Pi)}\Ep =\Ep e^{iS_{\rm free}(\overline{p})} .
 \label{Main}
\ee 
So, the $\Ep$ functions not only render the fermion propagator in external fields diagonal in momentum space, with a free form involving the quantum numbers induced by the field. 
These also allow to express the exact FW transformation in the presence of the fields in a free form.


For definitiveness, let us consider the case of electrons restricted to move in a plane under the influence of a static magnetic field pointing perpendicularly to their plane of motion. 
This can be viewed as an effective dimensional reduction of the ordinary problem neglecting the third spatial component, along which the magnetic field is aligned and thus does not play a relevant role that affects the following discussion. 

In order to fulfill the Clifford algebra $\{\gamma^\mu,\gamma^\nu\}=2g^{\mu\nu}$ in (2+1) dimensions, only  three $\gamma^\mu$-matrices are required.
The lowest dimensional representation of these matrices is $2\times 2$, and hence can be chosen to be proportional to the Pauli matrices. There are two inequivalent representations of the $\gamma^\mu$-matrices, namely 
\be
\gamma^{0} = \sigma_3, \quad
 \gamma^{1} =i\sigma_1, \quad 
 \gamma^{2} = i\sigma_2 ,
\label{primera}
\ee
 for which
$\gamma^\mu\gamma^\nu=g^{\mu\nu}-i\epsilon^{\mu\nu\lambda}\gamma_\lambda$;
and
\be
\tilde{\gamma}^{0} = \sigma_3, \quad 
\tilde{\gamma}^{1} = i\sigma_1, \quad 
\tilde{\gamma}^{2} =-i\sigma_2 ,
\label{segunda}
\ee
for which 
$\tilde{\gamma}^\mu\tilde{\gamma}^\nu=g^{\mu\nu}+i\epsilon^{\mu\nu\lambda}\tilde{\gamma}_\lambda$.
In what follows, we will adopt the first representation, Eq.~(\ref{primera}), but the discussion can be straightforwardly  extended to the second representation, Eq.~(\ref{segunda}).
Moreover, the external static inhomogeneous field is specified in a Landau-like gauge through the vector potential  $A^\mu = (0,0,W(x))$, such that $B(x) = W^\prime(x)$. 
In this case and recalling Eq.~(\ref{H_FW-Serie}),
$(\gammaPi)^2 = \mathbf{\Pi}^2 - e \sigma_3 W^\prime(x)$~\cite{Barducci:2009}.

Let us look at the problem of finding the  FW transformation
within our approach. 
To see the  usefulness of Eq.~(\ref{Main}), we
first apply the Hamiltonian in Eq.~(\ref{H_FW-Serie}) to the Ritus eigenfunctions ${\mathbb E}_p$,
\be
H_{\Pi}^{FW}  {\mathbb E}_p = \left( \gamma^0 \sqrt{(\Pit)^2 + m^2} \right) {\mathbb E}_p ,
\label{H_FW-Ep}
\ee
which has to be evaluated expanding the square-root operator in a power series of $(\Pit/m)^2$. This procedure leads to an expression in terms of the eigenvalues $k$ of the operator $(\Pit)^2$, namely
\be
(\Pit)^2 =-\gp^2 + \Pi_{0}^{2},
\label{PiTilde2}
\ee 
given through 
Eq.~(\ref{EigenvaluesEp2}) with $p^2 = \overline{p}^2$.
Since $p_0 = E_D$ are the eigenvalues of $\Pi_0 = i \partial_t$,
from the Dirac equation 
$p_{0}^{2}  = k + m^2$, thus $\sqrt{k}$ correspond to the energy eigenvalues of a particle on-shell.
From Eq.~(\ref{PiTilde2}), $\overline{p}^2 = p_{0}^{2} - k$, which can be fulfilled with the choice of $\overline{p}^\mu = (p_0, 0 , \sqrt{k})$, in accordance to our selection of gauge.
Hence, Eq.~(\ref{H_FW-Ep}) simplifies to 
\be
H_{\Pi}^{FW}  {\mathbb E}_p = {\mathbb E}_p \left( \gamma^0 \sqrt{E_{D}^{2} + m^2} \right) .
\label{H_FW-Ep-Eigenvalues}
\ee

On the other hand, notice that under the FW transformation, the Hamiltonian $H=\gamma^0 (\boldsymbol{\gamma \cdot \overline{p}} + m)$ on the r.h.s. of Eq.~(\ref{DiracEpFree}) transforms in a free form, as in Eq.~(\ref{H_FW-free}), but involving $\overline{\bf p}^2$ alone. Thus the FW transformed Hamiltonian~(\ref{DiracEpFree}) can be written directly:
\be
H_{\rm free}^{FW}  {\mathbb E}_p = 
     {\mathbb E}_p \left( \gamma^0 \sqrt{E_{D}^{2} + m^2} \right) .
\label{H_FW-Ep-Eigenvalues-Free}
\ee
The r.h.s. of this last equation precisely corresponds to the r.h.s. of Eq.~(\ref{H_FW-Ep-Eigenvalues}). This last was obtained transforming the Dirac Hamiltonian of Eq.~(\ref{DiracEp}) with a magnetic filed in the {\it usual} way. As comparison, with the aid of Eq.~(\ref{DiracSpinor}), the corresponding FW transformed Hamiltonian was obtained directly from a free one, Eq.~(\ref{DiracEpFree}), given in terms of the tri-momentum $\overline{p}^\mu$ which contains all the dynamics induced by the external magnetic field.
It is then straightforward to prove the relationship between the FW transformations $S(\Pi)$ and $S_{\rm free}(\overline{p})$ established by Eq.~(\ref{Main}) in terms of the Ritus eigenfunctions $\Ep$.


Summarizing, we have shown that the Ritus eigenfunctions provide a direct calculation of the exact FW transformation of the Dirac Hamiltonian in presence of external static electromagnetic fields.  In the Ritus basis, the FW  transformation can be expressed  in a closed form in terms of a free one which depends on the dynamical quantities induced by the fields.


\begin{acknowledgments}
The authors are indebted to Mat\'ias Moreno, Christian Schubert and Manuel Torres for valuable discussions and careful
reading of the manuscript. Support has been received from CIC-UMSNH under project 4.22. AR acknowledges support from SNI and
CONACyT grants under project 82230.
GM acknowledges support from DGAPA-UNAM grant under project PAPIIT IN118610.
\end{acknowledgments}


\end{document}